\documentclass[english,aps,prl,twocolumn,amsmath,amssymb,showpacs,superscriptaddress,notitlepage,longbibliography]{revtex4-1}
\usepackage{bm}
\usepackage{hhline}
\usepackage{bbold}
\usepackage{makecell}
\usepackage{amsmath}
\usepackage{amssymb}
\usepackage{mathdots}
\usepackage{tabularx}
\usepackage{graphicx}
\usepackage{hyperref}
\usepackage{bbding}
\usepackage{tikz}
\usepackage{bbding} 
\usepackage{braket}
\usepackage{bm}

\makeatother

\usepackage{babel}
\begin{document}
\renewcommand{\figurename}{Fig.}
\title{Transverse superconducting diode without parity and time-reversal violation}
 \author{Ruo-Peng Yu}

\author{Jin-Xin Hu}

   \author{Zi-Ting Sun}\thanks{Contact author: zsunaw@connect.ust.hk}
 \affiliation{Department of Physics, The Hong Kong University of Science and Technology, Clear Water Bay, Hong Kong SAR, China}
   
\affiliation{Center for Theoretical Condensed Matter Physics, The Hong Kong University of Science and Technology, Clear Water Bay, Hong Kong SAR, China}

	\date{\today}
	\begin{abstract}

The superconducting diode effect (SDE) is characterized by its nonreciprocal nature in critical supercurrents. However, realizing a longitudinal SDE typically requires simultaneous time-reversal ($\mathcal{T}$) and inversion ($\mathcal{P}$) symmetry breaking in the device, raising challenges in applications. In this Letter, we reveal that an off-axis direct-current bias applied to a planar anisotropic superconductor can convert intrinsic anisotropy into transverse nonreciprocity, generating ultra-tunable SDE without breaking either $\mathcal{P}$ or $\mathcal{T}$ symmetry. Using both Ginzburg-Landau theory and self-consistent mean-field calculations, we show that diode efficiency can be continuously tuned via bias current amplitude. Notably, when the injected bias current exceeds a critical threshold, the system is driven into a ``unidirectional superconductivity" regime, where transverse dissipationless currents are permitted in only one direction. Based on this mechanism, we propose the ``current-gated orthogonal superconducting transistor (CGOST)" and demonstrate its utility in tunable supercurrent range controllers and half-wave rectifiers. Our findings open new avenues for nonreciprocal superconducting electronics.

	\end{abstract}
	\maketitle
    
\emph{Introduction.}---Nonreciprocal charge transport in superconductors~\cite{nadeem2023superconducting,nagaosa2024nonreciprocal}, specifically the superconducting diode effect (SDE) where critical currents are asymmetric ($j_{c+}\neq-j_{c-}$), has recently attracted intense interest in both theory \cite{hu2007proposed,dolcini2015topological,chen2018asymmetric,misaki2021theory,yuan2022supercurrent,he2022phenomenological,daido2022intrinsic,daido2022superconducting,tanaka2022theory,davydova2022universal,zhang2022general,ilic2022theory,he2023supercurrent,hu2023josephson,lu2023tunable,li2023valley,he2023supercurrent,sun2024flat,zeng2025transverse,daido2025unidirectional,hu2025geometric,daido2025nonreciprocal,shaffer2025theories} and experiments \cite{ando2020observation,lin2022zero,narita2022field,jeon2022zero,pal2022josephson,bauriedl2022supercurrent,baumgartner2022supercurrent,diez2023symmetry,gupta2023gate,trahms2023diode,hou2023ubiquitous,zhang2025observation,nagata2025field} for its immense promise in rectification and superconducting electronics~\cite{daido2024rectification,castellani2025superconducting,ingla2025efficient}. However, existing implementations are typically realized in two-terminal geometries and thus face a fundamental bottleneck: they rely on static time-reversal symmetry ($\mathcal{T}$) breaking (e.g., external magnetic fields or intrinsic magnetic orders) and inversion symmetry ($\mathcal{P}$) breaking (e.g., noncentrosymmetric structures), which restricts functional diversity and complicates control and integration of the devices.

In this work, we propose a generic scheme to overcome these limitations by converting crystalline or pairing anisotropy into transverse nonreciprocity through a multi-terminal current-gated orthogonal superconducting transistor (CGOST). The device architecture is illustrated in Fig.~\ref{fig:fig1}(a), where an anisotropic superconductor with a planar two-fold symmetry is formed into a disk-like geometry with multiple terminals. A direct current (DC) bias $j_{\mathrm{bias}}$ is injected along a direction $\hat{\theta}$ deliberately misaligned with the principal axes of the crystal (dashed lines) through the control terminal of CGOST. Consequently, the transverse critical currents $j_{mc\pm}$, measured orthogonally to the bias direction via the function terminal, exhibit pronounced asymmetry (Fig.~\ref{fig:fig1}(b)), with neither $\mathcal{T}$ nor $\mathcal{P}$ symmetry breaking. Remarkably, when the bias amplitude exceeds its critical current, both upper and lower transverse critical currents acquire the same sign. In this regime, supercurrent can flow only in one transverse direction while being completely suppressed in the opposite direction—a phenomenon identified as unidirectional superconductivity (USC)~\cite{lin2022zero,daido2025unidirectional,daido2025nonreciprocal}.

\begin{figure}
		\centering
		\includegraphics[width=1\linewidth]{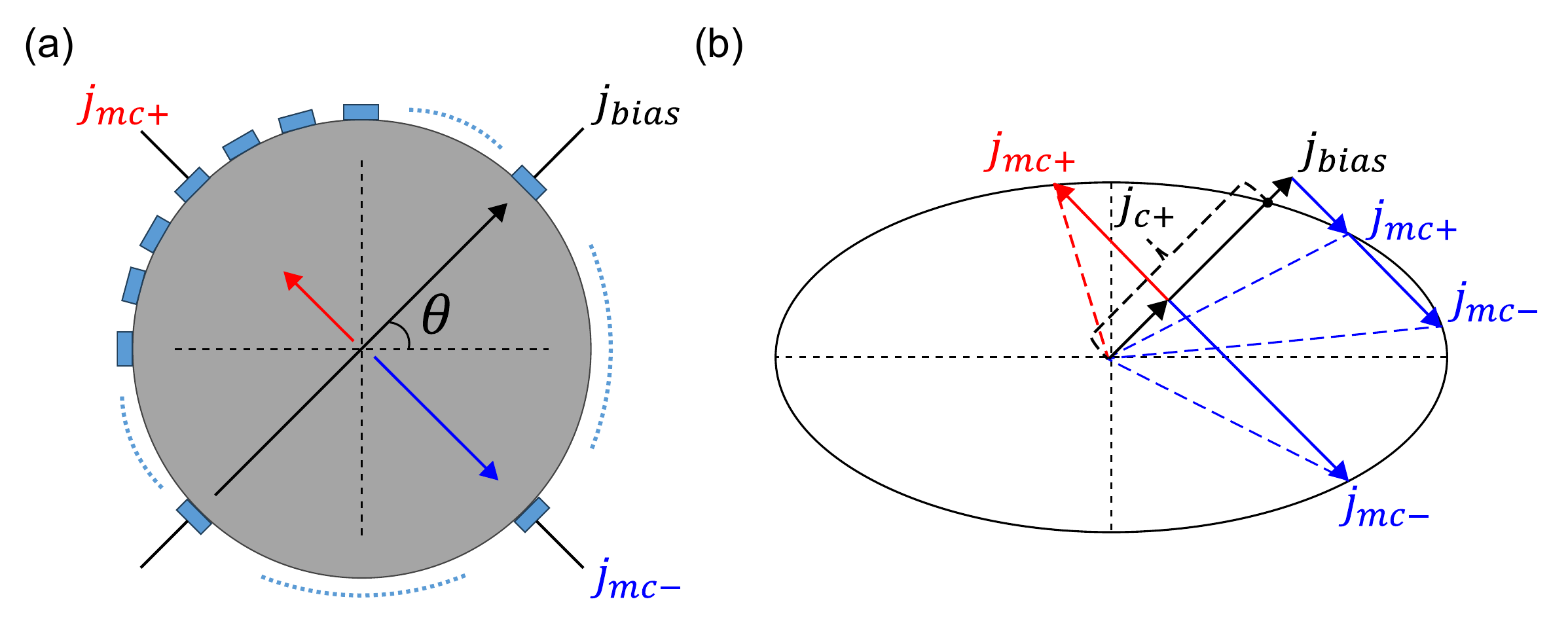}
		\caption {(a) Device schematic of CGOST. A bias current $j_{bias}$ is injected through the control terminal (black arrow) along a direction $\hat{\theta}$ relative to the crystal axes (dashed lines). Transverse critical currents, $j_{mc\pm}$, are measured at the functional terminal orthogonal to the bias and exhibits SDE. (b) Mechanism of anisotropy-transverse nonreciprocity conversion in a two-fold anisotropic superconductor. The elliptical envelope represents the distribution of critical currents. Red (blue) vectors indicate critical currents with positive (negative) transverse components. When $j_{bias}$ is non-zero, the transverse critical currents become asymmetric ($j_{mc+} \neq -j_{mc-}$). If $j_{bias}$ exceeds the critical current along the forward direction ($j_{c+}(\theta)$), both transverse critical currents acquire the same sign, enabling unidirectional superconductivity.}
		\label{fig:fig1}
\end{figure}

Combining phenomenological Ginzburg-Landau (GL) theory and self-consistent mean-field (MF) calculations, we report several key results: (i) CGOST is platform-agnostic, applicable to both normal-state anisotropy (anisotropic Fermi velocities in $s$-wave superconductors) and pairing-channel anisotropy (nematic $p$-wave pairing with isotropic Fermi surfaces); (ii) The mechanism naturally extends to high-temperature superconductors (high-$T_c$ SCs) with four-fold rotational symmetry, such as $d$-wave cuprates, significantly expanding operational temperature and current magnitude ranges; (iii) Two immediate applications based on CGOST are demonstrated, including a tunable supercurrent range controller~\cite{sun2025perfect} and the half-wave rectifier~\cite{castellani2025superconducting,ingla2025efficient} with adjustable rectification windows; and (iv) We provide design rules for optimizing device performance via simple bias angle selection. These findings establish CGOST as a versatile, multi-terminal building block for current-controlled superconducting electronics.

\emph{Ginzburg-Landau theory of anisotropic superconductors.}---Without loss of generality, we consider the phenomenological GL theory of a two-dimensional (2D) superconductor with two-fold rotational symmetry (the formalism for generic anisotropic superconductors can refer to \textbf{Supplemental Material (SM) Note}~\textbf{I}~\cite{supp}). This scenario may originate from nematicity in the normal state of conventional superconductors, as in Ising superconductors~\cite{lu2015evidence,xi2016ising} (like monolayer T$_{\rm{d}}$-MoTe$_{2}$ \cite{li2024topological}), and kagome superconductors AV$_3$Sb$_5$ (A=K, Rb, Cs)~\cite{zhao2021cascade,feng2021chiral,wang2024spin,feng2025odd}. Alternatively, it can arise from the anisotropic pairing, such as in nematic $p$-wave superconductors~\cite{sigrist1991phenomenological,sigrist2005introduction} (see \textbf{SM Note}~\textbf{II}~\cite{supp}), even with isotropic Fermi velocities.

First of all, the GL free energy is written as
\begin{equation}\label{free}
    F = \int d \boldsymbol{r}\{\frac{1}{2 m_x}|\partial_x\Psi|^2 + \frac{1}{2 m_y}|\partial_y\Psi|^2 + \alpha|\Psi|^2 + \beta|\Psi|^4\},
\end{equation}
where $m_x$ and $m_y$ are the effective mass along the $x$ and $y$-directions and $m_x \neq  m_y$ describes the anisotropy. $\alpha$ and $\beta > 0$ are GL coefficients. To describe the current-carrying superconducting state, we perform the Fourier transformation on the order parameter as $\Psi\equiv\Psi(\boldsymbol{r})=\frac{1}{\sqrt{\Omega}} \sum_{\boldsymbol{q}} \Psi_{\boldsymbol{q}} e^{i \boldsymbol{q} \cdot \boldsymbol{r}}$, and restrict the analysis to a single-$\boldsymbol{q}$ state selected by the injecting current. The free energy density for the center of mass momentum $\boldsymbol{q}$ state reads
\begin{equation}
    \mathcal{F}(\bm{q}) = \left( \frac{q_x^2}{2m_x} + \frac{q_y^2}{2m_y} + \alpha \right) |\Psi_{\bm{q}}|^2 + \beta |\Psi_{\bm{q}}|^4,
    \label{eq:3}
\end{equation}
where $\boldsymbol{q}=(q_x,q_y)$ satisfies $ {q_x^2}/{m_x} + {q_y^2}/{m_y} +2\alpha< 0$ in the superconducting regime. Then the corresponding supercurrent density can be obtained as:
\begin{equation}
         \bm{j}(\bm{q}) = - \frac{e}{\beta} \left[ \alpha + \frac{1}{2}\left( \frac{q_x^2}{m_x} + \frac{q_y^2}{m_y} \right) \right] \left( \frac{q_x}{m_x}\hat{x} + \frac{q_y}{m_y}\hat{y} \right).
         \label{eq:4}
\end{equation}
 It is worth mentioning that, although our discussion focuses on 2D superconductors, this method also describes the in-plane supercurrents of a three-dimensional superconductor with planar anisotropy.
 
To evaluate the depairing currents $j_{c\pm}(\theta)$ along an arbitrary direction $\hat{\theta}=(\cos\theta,\sin\theta)$ (for simplicity, we set $0 < \theta < \pi/2$), we firstly solve the momenta $\bm{q}_c(\theta)$ corresponding to the critical current, yielding
\begin{equation}
        \bm{q}_c(\theta) = \pm \frac{\sqrt{2|\alpha|}\left(m_x\cos\theta,m_y\sin\theta\right) }{\sqrt{3\left(  m_x\cos^2\theta + {m_y\sin^2\theta }\right)}}.
         \label{eq:5}
\end{equation}
Thus the domain $\mathcal{D}$ of all physical current-carrying states $\bm{q}$ is bounded by the elliptical envelope of $\bm{q}_c(\theta)$ (Fig.~\ref{fig:fig1}(b)): ${q_x^2}/{m_x} + {q_y^2}/{m_y} < 2|\alpha|/3$, which is consistent with the previous constraint ${q_x^2}/{m_x} + {q_y^2}/{m_y} < 2|\alpha|$ of $\boldsymbol{q}$. Substituting Eq.~\eqref{eq:5} into Eq.~\eqref{eq:4}, we have
\begin{equation}
        j_{c\pm}(\theta) = \pm \frac{e}{\beta} \sqrt{\frac{8|\alpha|^3}{27\left(  m_x\cos^2\theta + m_y{\sin^2\theta }\right)}}.
        \label{eq:jcpm}
\end{equation}
Here, $j_{c+}(\theta)=-j_{c-}(\theta)$ in the longitudinal direction is ensured by $\mathcal{T}$ or $\mathcal{P}$ symmetry. In this way, the anisotropy degree $m_x/m_y$ can be easily obtained from the experimental measurement of the critical current by $m_x/m_y=\left(j_{c\pm}(\pi/2)/j_{c\pm}(0)\right)^2$. For $j_{c\pm}(\pi/2)$ and $j_{c\pm}(0)$, it has been reported that one can be as large as twice the value of another~\cite{hecher2018direct,lyatti2024plane}.

\begin{figure}[t]
		\centering
		\includegraphics[width=1\linewidth]{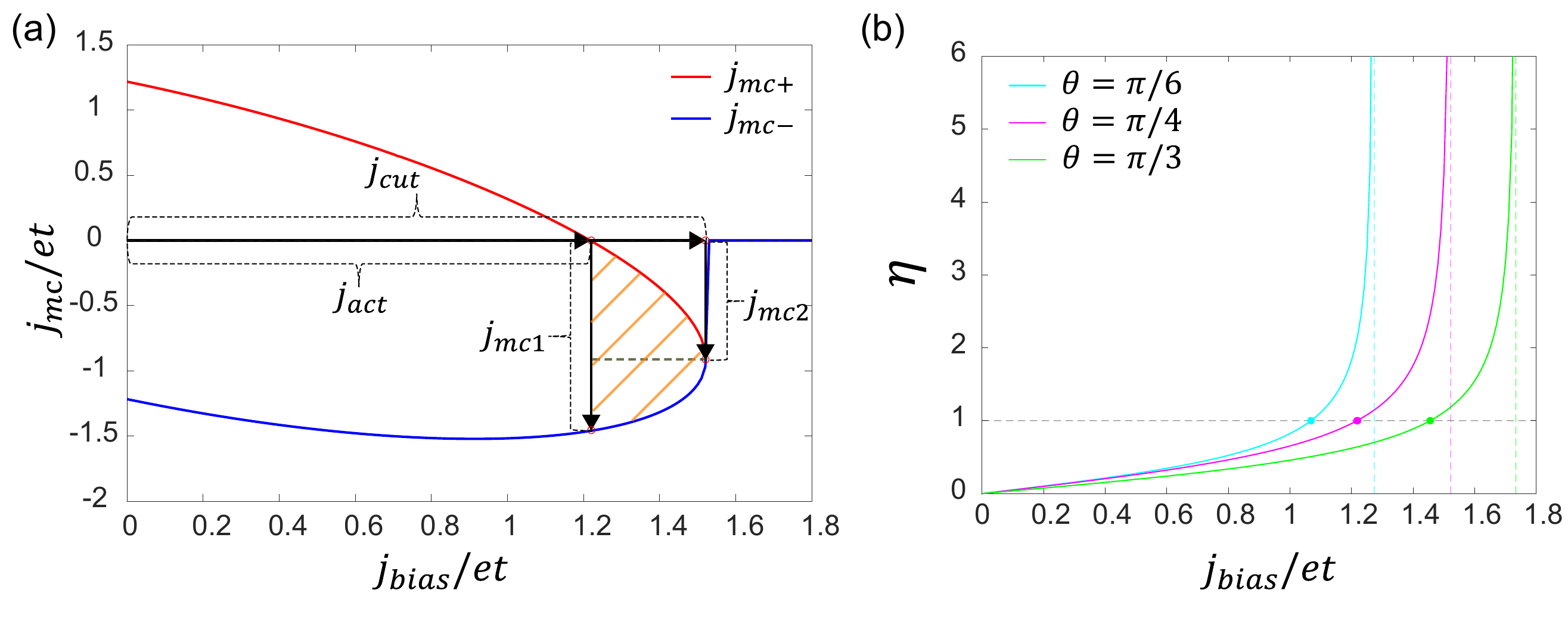}
		\caption{Evolution of transverse critical currents and diode efficiency: (a) The evolution of $j_{mc+}$ and $j_{mc-}$ (labeled as red and blue lines respectively) with increasing $j_{bias}$, $j_{mc+}$ becomes negative when $j_{bias} > j_{act}=1.217 et$, and the superconducting region vanishes when $j_{bias} >j_{cut} = 1.522 et$. Here we set $\theta = \pi/4$. (b) The diode efficiency $\eta(j_{bias})$ for three different $\theta$. Parameters used here: $\alpha = -0.5 t, \beta = 0.1 t, m_x = 4/t, m_y = 1/t$. $t$ is an arbitrary unit.}
		\label{fig:fig2}
\end{figure}

\emph{Transverse superconducting diode effect from direct current bias.}---Following our proposal in Fig.~\ref{fig:fig1}, now we apply a DC bias $j_{bias}$ in the $\hat{\theta}$ direction and measure the critical current along the transverse direction $\hat{\tau}=(\sin\theta,-\cos\theta)$. Consequently, we rewrite the total supercurrent in the polar system as $\bm{j}(\bm{q})=j_{\parallel}(\bm{q})\hat{\theta}+j_{m}(\bm{q})\hat{\tau}$, fix $j_{\parallel}(\bm{q})=j_{bias}$, and ask for the critical value of $j_{m}(\bm{q})$.

In Fig.~\ref{fig:fig2}(a) we show how the measured critical currents $j_{mc\pm}$ evolve as $j_{bias}$ increases. When $j_{bias} = 0$, we have $j_{mc+} = -j_{mc-}$. But when $j_{bias} \neq 0$ we immediately have $j_{mc+}(j_{bias}) \neq -j_{mc-}(j_{bias})$, which means that SDE can be observed in the transverse direction when the bias current $j_{bias}$ is applied. Despite this, global relations such as $j_{mc+}(j_{bias}) = -j_{mc-}(-j_{bias})$ remain enforced by overall $\mathcal{T}$ or $\mathcal{P}$ symmetry, which corresponds to $j_{c+} = -j_{c-}$ in a purely two-terminal configuration~\cite{nadeem2023superconducting,nagaosa2024nonreciprocal,shaffer2025theories}. The physical reason here is that, $\mathcal{T}$ and $\mathcal{P}$ symmetry is preserved on the space of all current carrying states of the system, but is broken ``kinematically" on the restricted space of $\bm{q}$ with the constraint $j_{\parallel}(\bm{q})=j_{bias}$, similar to the Doppler shift in superconductors~\cite{volovik1993superconductivity}. More detailed discussions of the conditions for bias-induced SDE can be found in \textbf{SM Note}~\textbf{III}~\cite{supp}.

A standard measurement of the nonreciprocity of SDE is the diode efficiency, which is defined as $\eta = \left|j_{mc+}+j_{mc-}|/|j_{mc+}-j_{mc-}\right|$~\cite{yuan2022supercurrent,daido2022intrinsic,he2022phenomenological}. Crucially, the diode efficiency $\eta$ is continuously tunable by the injected bias $j_{bias}$ ($\eta$ increases monotonically with $j_{bias}$ as presented in Fig.~\ref{fig:fig2}(b)), allowing transistor-like control. To evaluate the control ability of CGOST, we introduce the differential gain $g_m = \partial \eta / \partial j_{bias}$ as a counterpart to the transconductance of conventional semiconductor transistors~\cite{sze2008semiconductor}.

For the small $j_{bias}$ limit, $\eta$ is approximately linear in $j_{bias}$ as $\eta\approx g_m(\theta)j_{bias}$, and $g_m(\theta)$ reads (for derivation, see \textbf{SM Note}~\textbf{IV}~\cite{supp}):
\begin{equation}
g_m(\theta) =  \frac{\beta}{e}\left(\frac{3}{2|\alpha|}\right)^{3/2}\frac{|(m_y-m_x)\sin\theta\cos\theta|}{\sqrt{m_x\sin^2\theta+m_y\cos^2\theta}}.
\label{efficiency_small_limit}
\end{equation}
We can see that $\eta\neq0$ as long as $m_x\neq m_y$ and $\hat{\theta}$ is deviated from high symmetry axes, which makes $\sin\theta$ or $\cos\theta$ vanish. At this regime, we can seek the maximal value of $g_m(\theta)$ at $\tan \theta = \sqrt[4]{m_y/m_x}$ and
\begin{equation}
    \text{max}(g_m(\theta)) = \frac{\beta}{e}\left(\frac{3}{2|\alpha|}\right)^{3/2}\left|\sqrt{m_y}-\sqrt{m_x}\right|.
\end{equation}
We point out that this previously overlooked phenomenon can be regarded as an analogue of the crystal Hall effect \cite{vsmejkal2020crystal} in anisotropic superconductors, where an off-axis electric field can induce a Hall current response in anisotropic metals even in the presence of both $\mathcal{T}$ and $\mathcal{P}$ symmetries.

\emph{Unidirectional regime.}---Perhaps more surprisingly, when $j_{bias}$ exceeds its critical value (we denote as $j_{act}$, the activation current of USC), we have $j_{mc-}<j_{mc+}<0$, i.e., the USC, which had only been achieved in non-equilibrium states~\cite{daido2025unidirectional,shaffer2025josephson}. It emerges in the transverse direction (the orange region in Fig.~\ref{fig:fig2}(a)) as a fully equilibrium state. In \textbf{SM Note}~\textbf{III}~\cite{supp}, we have shown $j_{act}=j_{c+}(\theta)$. Here, the system firstly transitions to a normal state if a bias $j_{bias}>j_{act}$ is applied. Subsequently, by driving a transverse current $j_m$ while keeping $j_{bias}$ fixed, the system will reenter a superconducting state when $j_m<j_{mc+}$. However, the superconductivity is lost again when $j_m$ exceeds another critical threshold $j_{mc-}$. If we further increase the bias current beyond the cut-off current $j_{cut}$ (Fig.~\ref{fig:fig2}(a)), the superconducting region vanishes. 

The diode efficiency $\eta$ also shows these features as in Fig.~\ref{fig:fig2}(b), $\eta>1$ when $j_{bias}>j_{act}$ and finally gets divergent when $j_{bias}=j_{cut}$. Different from the $j_{bias}<j_{act}$ region with approximately $\eta\propto j_{bias}$ and constant $g_m$, $\eta$ is highly nonlinear dependent on $j_{bias}$ and so is $g_m$ in the USC region, implying a significant gain of CGOST (see \textbf{SM Note}~\textbf{IV}~\cite{supp} for more details).

To acquire a large USC operating range with a small activation current in CGOST, we introduce the USC quality factor $Q_{USC} = 1-j_{act}/j_{cut}$ and expect a larger value of $Q_{USC}$. As derived in \textbf{SM Note}~\textbf{IV}~\cite{supp}, the cut-off current $j_{cut}$ is given by:
\begin{equation}
    j_{cut} = \frac{e}{\beta}\left(\frac{2|\alpha|}{3}\right)^{3/2}\sqrt{\frac{\cos^2\theta}{m_x}+\frac{\sin^2\theta}{m_y}}.
\end{equation}
So the USC quality factor $Q_{USC}(\theta)$ is
\begin{equation}
1 - \sqrt{\frac{m_xm_y}{(m_x\cos^2\theta+m_y\sin^2\theta)(m_y\cos^2\theta+m_x\sin^2\theta)}}. 
\end{equation}
Notably, the maximum of $Q_{USC}(\theta)$ is always located at $\theta = \pi/4$, which is independent of the effective masses $m_x$ and $m_y$, and:
\begin{equation}
    \text{max}(Q_{USC}) = Q_{USC}(\pi/4) = 1-\frac{2\sqrt{m_xm_y}}{m_x+m_y}.
    \label{Q_USC_Max}
\end{equation}
We note that $\text{max}(Q_{USC})$ reveals a simple relationship to the degree of anisotropy $m_x/m_y$ of the superconductor. These simple rules guide the design of CGOST.

\emph{Self-consistent mean-field theory.}---To supplement the analysis based on our GL theory near $T_c$ , we perform numerical calculations using self-consistent MF theory near zero temperature for a square lattice model. The normal-state dispersion is given by $\xi_{\bm{k}} = -2t_x \cos(k_x) - 2t_y \cos(k_y) - \mu$, where $t_x$ and $t_y$ are the hopping amplitudes along the $x$ and $y$ directions, and we set the lattice constant $a=1$. We mainly consider two cases: (i) a spin-singlet $s$-wave superconductor with anisotropic hopping ($t_x \neq t_y$), and (ii) a nematic spin-triplet $p$-wave superconductor with isotropic hopping. In principle, both cases are captured by the model in Eq.~\eqref{free}.

In this formalism (for details, see \textbf{SM Note}~\textbf{V}~\cite{supp}), the current-flowing superconducting state can be captured by the finite-momentum pairing order parameter matrix $\hat{\Delta}(\bm{k},\bm{q})$ with a center-of-mass momentum $\bm{q}$~\cite{daido2022intrinsic,daido2022superconducting,sun2024flat}. For the spin-singlet $s$-wave pairing, $\hat{\Delta}(\bm{k},\bm{q})=\Delta(\bm{q})i\sigma_y$ and for the nematic spin-triplet $p$-wave pairing, $\hat{\Delta}(\bm{k},\bm{q}) = \Delta(\bm{q})\sin(k_x)\sigma_x$. $\Delta(\bm{q})$ can be determined by the self-consistent gap equation 
\begin{equation}
\label{eq:eq_pair_self}
    \Delta(\bm{q})=-\frac{U}{N_c} \sum_{\bm{k}}\left\langle\hat{c}_{-\bm{k} \downarrow} \hat{c}_{ \bm{k}+\bm{q} \uparrow}\right\rangle u_{\bm{k}},
\end{equation}
where the form factor $u_{\bm{k}}=1$ for the $s$-wave pairing and $u_{\bm{k}}=\sin{k_x}$ for the nematic $p$-wave pairing. Then we can obtain the minimized Helmholtz free energy density $\mathcal{F}_m(\bm{q})$ in terms of solved $\hat{\Delta}(\bm{k},\bm{q})$, and the supercurrent can be evaluated by $\vec{j}(\bm{q}) = 2e \nabla_{\bm{q}} \mathcal{F}_m(\bm{q})$.

The schematics and corresponding numerical results for these two cases are summarized in \textbf{SM Note V}~\cite{supp}, where the curves share similar features compared to GL results in Fig.~\ref{fig:fig2}(a). The SDE can occur as long as $j_{bias}>0$, and the USC can be realized when $j_{bias} \geq j_{c+}(\theta)$. A more quantitative comparison can also be found in \textbf{SM Note V}~\cite{supp}. Thus, we numerically verified that the anisotropies of both the Fermi velocity and the pairing symmetry can lead to bias-induced SDE and USC, which aligns with the predictions from GL theory.

\emph{high-$T_c$ superconductor platform.}---Up to now, we have focused on CGOST based on superconductors with two-fold anisotropy. Here, we point out that the high-$T_c$ SC platform is naturally compatible with our proposal, employing their four-fold pairing anisotropy. The primary advantage of this way stems from the ability of high-$T_c$ SC to operate at more economically viable temperatures and a sufficiently wide current operating range due to its large critical currents~\cite{orenstein2000advances,chen2016high}.

We consider the $d_{x^2-y^2}$-wave pairing case for instance, which transforms under the $B_{1g}$ irreducible representation of the tetragonal point group $D_{4h}$~\cite{sigrist1991phenomenological,sigrist2005introduction}. Standing as a typical pairing symmetry in cuprate superconductors~\cite{sigrist1991phenomenological,orenstein2000advances,sigrist2005introduction}, the corresponding GL free energy is
\begin{equation}
    \begin{aligned}
             F &= \int d \boldsymbol{r}\{\alpha_0|\Psi|^2+A_{1}^d \left(|\partial_x\Psi|^2+|\partial_y\Psi|^2\right) \\ &
            +A_{2}^{d}|(\partial^2_x+\partial^2_y)\Psi|^2+ A_{3}^{d}|(\partial^2_x-\partial^2_y)\Psi|^2 + \beta_0|\Psi|^4.
    \end{aligned}
\end{equation}
In momentum space, the free energy density is in the form of $\mathcal{F} = \alpha(\bm{q}) |\Psi_{\bm{q}}|^2 + \beta |\Psi_{\bm{q}}|^4$ with $\alpha(\bm{q})=\alpha_0 + K_{1}^{d}\left(q_x^2 +q_y^2\right) + K_{2}^{d}\left(q_x^4 +q_y^4\right) + K_{3}^{d} q_x^2q_y^2$. The coefficients are redefined as: $K_{1}^{d} = A_{1}^{d}$, $K_{2}^{d}=A_{2}^{d}+A_{3}^{d}$, $K_{3}^{d} = 2A_{2}^{d} - 2A_{3}^{d}$. 
Similarly, the boundary of the physical domain can be determined as in \textbf{SM Note}~\textbf{I}~\cite{supp}, which is shown in Fig.~\ref{fig:fig3}(a) as the black curve, and the envelope of critical currents $j_{c\pm}(\theta)$ is shown in Fig.~\ref{fig:fig3}(a) as the blue curve, both have apparent four-fold anisotropy.

\begin{figure}[t]
		\centering
		\includegraphics[width=1\linewidth]{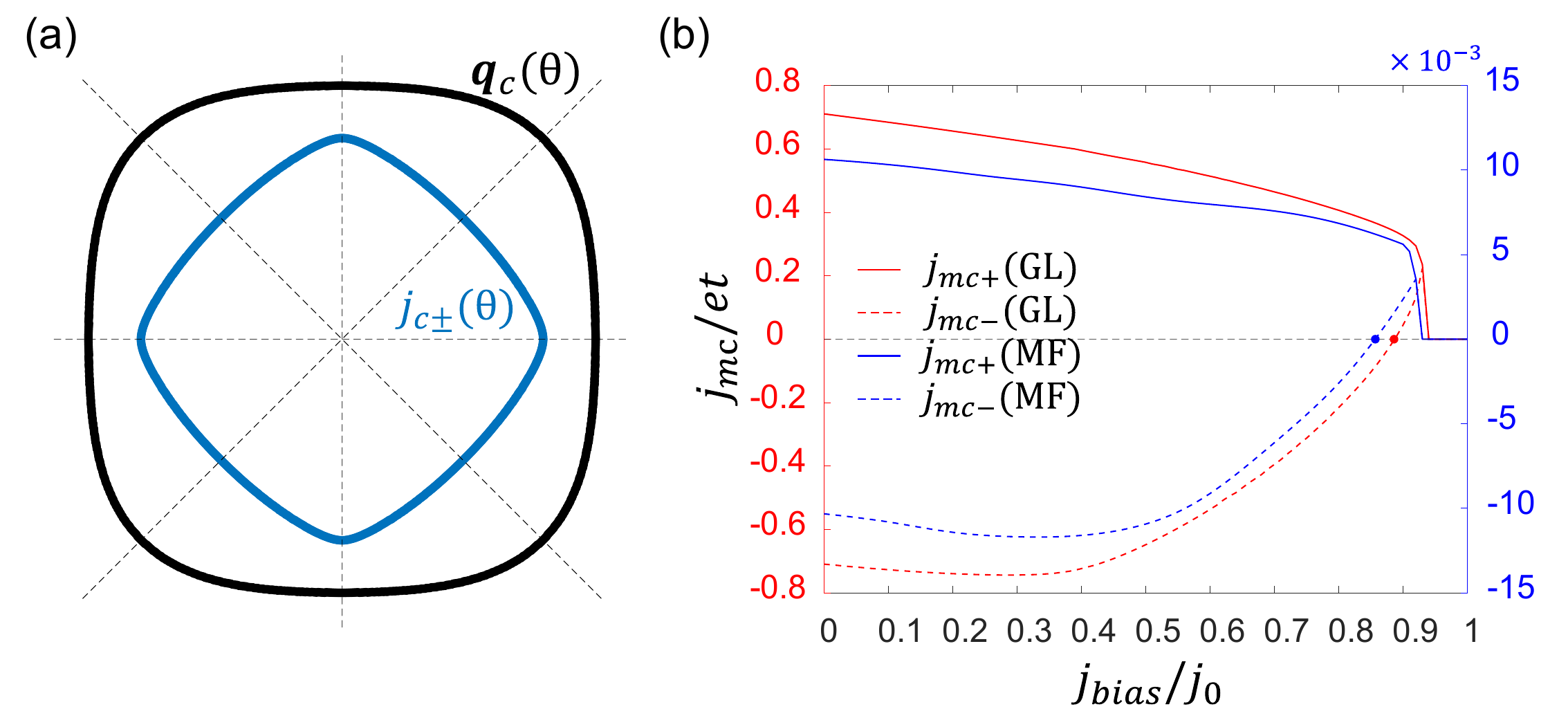}
		\caption{Analysis of $d$-wave superconductors: (a) The black curve shows the envelope of the critical momenta $\bm{q}_c(\theta)$. The blue curve shows the envelope of the critical current $j_{c\pm}(\theta)$. Both are derived from the GL analysis of the $d_{x^2-y^2}$-wave superconductor. (b) The evolution of $j_{mc+}$ (solid lines) and $j_{mc-}$ (dashed lines) from the GL analysis (in red) and MF calculations (in blue), $j_{bias}$ is rescaled to $[0,1]$ in arbitrary units $j_0$.  Parameter used in GL analysis: $\theta = \pi/8, \alpha_0 = -t, \beta_0 = 2t, K_{1}^{d} = 0.5t, K_{2}^{d} = 4t, K_{3}^{d} = t$, then $j_{act} = j_{c+}(\theta) = 0.7096et, j_{act}/j_{0}=0.8870$ with $j_{0}=0.8et$, and $Q_{USC} = 5.4\%$. Parameter used in MF calculation: $\theta = \pi/8, t_x = t_y = t, \mu = 2.5t, U = 3t$, then $j_{act} = j_{c+}(\theta) = 0.0112et$, $j_{act}/j_{0}=0.8615$ with $j_{0}=0.013et$ and $Q_{USC} = 7.6\%$.}
        \label{fig:fig3}
\end{figure}

Accompanying the GL analysis, we also performed numerical MF calculations (for method, refer \textbf{SM Note}~\textbf{V}~\cite{supp}) with spin-singlet $d_{x^2-y^2}$-wave pairing $\hat{\Delta}(\bm{k},\bm{q})=\Delta(\bm{q})\left(\cos{k_x} - \cos{k_y}\right)i\sigma_y$ and using $u_{\bm{k}}=\cos{k_x} - \cos{k_y}$ in Eq.~\eqref{eq:eq_pair_self}. Fig.~\ref{fig:fig3}(b) shows a comparison between the results of the GL theory (red lines) and the numerical calculations (blue lines), which are in qualitative agreement. This indicates that our phenomenological model and the chosen parameters well capture the essential features of the $d_{x^2-y^2}$-wave pairing, establishing a foundation for its use in CGOST.

\emph{Bias-controlled supercurrent range controller and rectifier.}---Based on previous discussions, we present two examples of how to design novel superconducting electronic devices using the CGOST, i.e., a DC bias-controlled supercurrent range controller and a superconducting half-wave rectifier.

For the supercurrent range controller~\cite{sun2025perfect}, we tune the CGOST to the USC region, then the nonvanishing supercurrent in the functional terminal is restricted in the narrow, adjustable region between $\left[j_{mc-},j_{mc+}\right]$ excludes zero current (the region highlighted in orange in Fig.~\ref{fig:fig2}(a)). Initially, we apply $j_{bias}=j_{act}$ and the we have a maximal supercurrent range $\left[j_{mc1},0\right]$. By increasing $j_{bias}$ from $j_{act}$ to $j_{cut}$, the controlling range can be narrowed from $\left[j_{mc1},0\right]$, finally to a single value at $j_{mc2}$. Thus $j_{mc1}$ and $j_{mc2}$ are key parameters for the supercurrent range controller.

Roughly speaking, the center of the controlling range is at $j_{mc2}$ (see the grey dashed line in Fig.~\ref{fig:fig2}(a)), whose tunability relies on
\begin{equation}
    j_{mc2} = \frac{e}{\beta}\left(\frac{2|\alpha|}{3}\right)^{3/2} \frac{\left(m_y-m_x\right)\sin\theta \cos\theta}{\sqrt{\left(m_x \sin ^2 \theta+m_y \cos ^2 \theta\right) m_x m_y}}.
\end{equation}
And we can see that the anisotropy and bias direction play decisive roles here.

Then we discuss the design for the half-wave rectifier. As schematically shown in Fig.~\ref{fig:fig4}(a), in the circuit, the input AC current $I_{in} = I_0 \sin(\omega t)$ is to be rectified in the functional terminal of CGOST, where $I_0$ is the amplitude and $\omega$ is the frequency. Here, CGOST is approximately in an equilibrium state at any time $t$ as long as the frequency is not too large. We note that the rectification window of this device can be tuned in the same way as the supercurrent range controller. To optimize the use of the original current waveform, the applied DC bias should be set to $j_{bias}=j_{act}$ to ensure an ideal rectification window $\left[j_{mc1},0\right]$, and the amplitude of the input current satisfies $I_{0} \leq |j_{mc1}|$. We denote $R_d$ as the resistance of the CGOST, and Fig.~\ref{fig:fig4}(b) shows how it evolves under the input current $I_{in}$. When $I_{in} \leq  0$ the CGOST is in the superconducting regime ($R_d = 0$), and when $I_{in} > 0$ it is in the normal state ($R_d \approx R_0$). Fig.~\ref{fig:fig4}(c) shows the output voltage $V_{out}$ and the output current $I_{out}$, realized a half-wave manner of rectification.

We define $Q_{rec} = |{j_{mc1}}/{j_{act}}|$ to measure the quality of the rectifier made from CGOST. A larger $Q_{rec}$ implies a wider rectification window with respect to a smaller activation bias current. In terms of Eq.~\eqref{eq:3}, we can show that (see the details in \textbf{SM Note}~\textbf{IV}~\cite{supp}):
\begin{equation}
        Q_{rec}(\theta) = \frac{2|m_x-m_y|}{m_x\tan\theta+m_y\cot\theta},
\end{equation}
which provides a simple angle-dependent design rule for the device optimization. The maximum of $Q_{rec}(\theta)$ is located at $\tan\theta = \sqrt{m_y/m_x}$, and $\text{max}(Q_{rec}) =  \frac{|m_x-m_y|}{\sqrt{m_ym_x}}$, implying that larger anisotropy gives rise to better rectifier, as shown in Fig.~\ref{fig:fig4}(d).

\begin{figure}[t]
		\centering
		\includegraphics[width=1\linewidth]{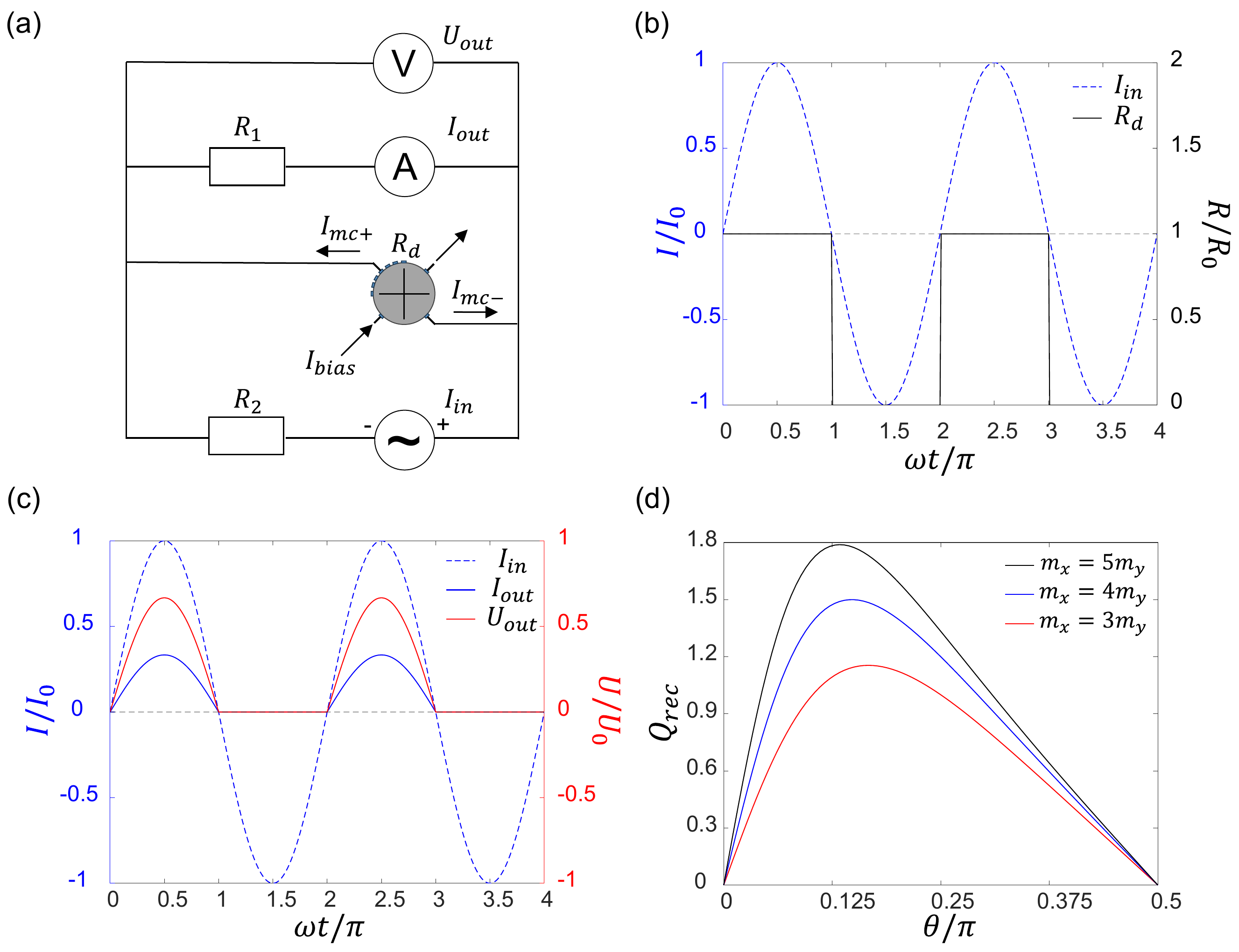}
		\caption{(a) Circuit schematic of the bias-controlled supercurrent rectifier. The input AC current $I_{in} = I_0 \sin(\omega t)$, the load resistance $R_1 = R_2 = 2R_0$. $R_d$ is the resistance of the CGOST. The rectifier is set to its working regime, so the bias current $j_{bias}=j_{act}$ and the amplitude of the input current $I_{0} \leq |j_{mc1}|$. (b) The evolution of $R_d$ (black solid line) with the input AC current $I_{in}$ (blue dashed line). (c) The evolution of the output voltage $U_{out}$ (red solid line) and the output current $I_{out}$ (blue solid line) with the input AC current $I_{in}$ (blue dashed line), $U_0 = I_0R_0$. (d) The quality factor $Q_{rec}(\theta)$ of different effective mass $m_x$ and $m_y$.   }
		\label{fig:fig4}
\end{figure}

\emph{Conclusion and discussion.}---In summary, we have demonstrated that a DC bias can generally induce a tunable transverse SDE and USC in anisotropic superconductors. Our proposal harnesses intrinsic material anisotropy and the underdeveloped multi-terminal design~\cite{gupta2023gate,zeng2025transverse} to achieve tunable nonreciprocal superconducting transport without explicit $\mathcal{T}$ and $\mathcal{P}$ symmetry breaking. We anticipate that our design protocol will significantly improve the accessibility of nonreciprocal phenomena in superconducting systems, and provide a new paradigm for designing novel nonreciprocal superconducting devices.

Throughout this work, we assume the $j_{bias}$ is injected from external terminals. However, it can also be created from the Meissner effect as screening currents \cite{tinkham2004introduction}, without extra terminals. This observation suggests different designs for CGOST, leaving room for future exploration.

\emph{Acknowledgements}---The authors thank Ying-Ming Xie, Wen-Yu He, James Jun He, K. T. Law, Akito Daido and Su-Yang Xu for inspiring discussions. This work is supported by the School of Science, the Hong Kong University of Science and Technology.

\bibliographystyle{apsrev4-2}
\bibliography{Manuscript}

\clearpage
\onecolumngrid
\begin{center}
\textbf{\large Supplemental Material for ``Transverse superconducting diode without parity and time-reversal violation''}\\[.2cm]
Ruo-Peng Yu,$^{1,2}$   Jin-Xin Hu,$^{1,2}$  Zi-Ting Sun$^{1,2}$  \\[.1cm]
{\itshape ${}^1$Department of Physics, The Hong Kong University of Science and Technology, Clear Water Bay, Hong Kong SAR, China}

{\itshape ${}^2$Center for Theoretical Condensed Matter Physics, The Hong Kong University of Science and Technology, Clear Water Bay, Hong Kong SAR, China}
\end{center}

\maketitle

\setcounter{equation}{0}
\setcounter{section}{0}
\setcounter{figure}{0}
\setcounter{table}{0}
\setcounter{page}{1}
\renewcommand{\theequation}{S\arabic{equation}}

\renewcommand{\thefigure}{S\arabic{figure}}
\renewcommand{\thetable}{S\arabic{table}}
\renewcommand{\tablename}{Supplementary Table}

\renewcommand{\bibnumfmt}[1]{[S#1]}
\renewcommand{\citenumfont}[1]{#1}

\makeatletter
\maketitle


\section{\bf{\uppercase\expandafter{I. GL analysis of the generic anisotropic superconductor carrying supercurrent}}}

We start from a generic single-$\bm{q}$ free energy density written as:
\begin{equation}
    \mathcal{F} = \alpha(\bm{q}) |\Psi_{\bm{q}}|^2 + \beta |\Psi_{\bm{q}}|^4,
    \label{app:eqs1}
\end{equation}
where $\beta > 0$ and $ \alpha(\bm{q})< 0$ denote the superconducting regime. For simplicity, we neglect the $\bm{q}$-dependence in the coefficient of the quartic term.

To calculate the current density, $|\Psi_{\bm{q}}|^2$ can be determined by minimizing the free energy density: $\frac{\partial \mathcal{F}}{\partial |\Psi_{\bm{q}}|^2} = 0$, so $|\Psi_{\bm{q}}|^2 = -\frac{\alpha(\bm{q})}{2\beta}$ and the minimized free energy density is $\mathcal{F}_m=-\frac{\alpha^2(\bm{q})}{4\beta}$, then the current density is given by:
\begin{equation}\label{app:eqs2}
        \bm{j}(\bm{q}) = 2e \nabla_{\bm{q}} \mathcal{F}_m= - \frac{e\alpha(\bm{q})}{\beta} \left(\alpha_x^{\prime} \hat{x} + \alpha_y^{\prime} \hat{y} \right),
\end{equation}
where $\alpha_i^{\prime} = {\partial  \alpha(\bm{q})}/{\partial q_i}$.

Then we consider the critical current along the $\hat{\theta}=(\cos\theta,\sin\theta)$ direction $j_{c\pm}(\theta)$, which is equivalent to find the extreme values of $|\bm{j}(\bm{q})|$ along $\hat{\theta}$, in this case, the components of $\bm{j}(\bm{q})$ satisfy:
\begin{equation}
        \frac{j_y(\bm{q})}{j_x(\bm{q})} = \frac{ \alpha_x^{\prime}}{\alpha_y^{\prime}} = \tan\theta.
\end{equation}

To find the extreme values of $|\bm{j}(\bm{q})|$ with such a constraint, we can use the method of Lagrange multipliers, i.e., $\mathcal{L} = |\bm{j}(\bm{q})| + \lambda(\alpha_x^{\prime}-\alpha_y^{\prime}\tan\theta)$.
The extreme values can be found by solving the following equations simultaneously:
\begin{equation}
\left\{\begin{aligned}
& \alpha_x^{\prime}=\alpha_y^{\prime}\tan\theta \\
& \frac{\partial\mathcal{L}}{\partial q_x} =\frac{\partial\mathcal{L}}{\partial q_y} =0 \\
\end{aligned}
\label{app:setofeq1}
\right. ,
\end{equation}
so that we can show all the local maxima $\bm{q}_c(\theta)$ of $|\bm{j}(\bm{q})|$ are located at the curve:
\begin{equation}
    \begin{aligned}
            2\alpha_x^{\prime}\alpha_y^{\prime}\alpha_{xy}^{\prime\prime} & = \left(\alpha_x^{\prime} \right)^2 \alpha_{yy}^{\prime\prime} + \left( \alpha_y^{\prime}\right)^2 \alpha_{xx}^{\prime\prime} \\
            & + \alpha(\bm{q})\left( \alpha_{xx}^{\prime\prime} \alpha_{yy}^{\prime\prime} -(\alpha_{xy}^{\prime\prime})^2\right),
    \end{aligned}
    \label{app:boundary1}
\end{equation}
where $ \alpha_{ij}^{\prime\prime} = \partial^2 \alpha(\bm{q})/\partial q_i \partial q_j$. Then we can set the boundary of the domain for momentum $\bm{q} \in\mathcal{D}$ according to Eq.~\eqref{app:boundary1}, to contain all physical points, and the critical current can be evaluated with the solutions of Eq.~\eqref{app:setofeq1} by $j_{c\pm}(\theta) = \pm |\bm{j}(\bm{q}_c(\theta))|$.

Then we assume that the bias current is applied along the $\hat{\theta}$ direction, then
the radical component $j_{\parallel}$ is:
\begin{equation}
    \begin{aligned}
        j_{\parallel} & = 2e \left(\cos\theta\hat{x} + \sin\theta\hat{y}\right) \cdot \nabla_{\bm{q}} \mathcal{F}_m \\
        & = - \frac{e\alpha(\bm{q})}{\beta} \left(  \alpha_x^{\prime}\cos\theta + \alpha_y^{\prime}\sin\theta \right),
    \end{aligned}
    \label{app:radicalj}
\end{equation}
and the transverse component $j_{m}$ is:
\begin{equation}
    \begin{aligned}
        j_{m} & = 2e \left(\sin\theta\hat{x} - \cos\theta\hat{y}\right) \cdot \nabla_{\bm{q}} \mathcal{F}_m \\
        & = - \frac{e\alpha(\bm{q})}{\beta} \left(  \alpha_x^{\prime}\sin\theta - \alpha_y^{\prime}\cos\theta \right).
    \end{aligned}
    \label{app:transversej}
\end{equation}

\section{\bf{\uppercase\expandafter{II. GL theory of a nematic $p$-wave superconductor}}}

We start from considering the GL formalism of a spin-triplet $p$-wave superconductor with tetragonal point group symmetry $D_{4h}$. The order parameter can be described by a two-component $\vec{d}$-vector: $\vec{d}(\bm{k})=\eta_x \hat{z} k_x+\eta_y \hat{z} k_y$, in the gap function $\hat{\Delta}(\bm{k})=i\vec{d}(\bm{k})\cdot\vec{\sigma}\sigma_y$. It transforms under the $E_{u}$ irreducible representation, and the generic GL free energy is given by \cite{sigrist1991phenomenological}:
\begin{equation}
\begin{aligned}
& F = \int d \boldsymbol{r}\{ a|\vec{\eta}|^2+b_1|\vec{\eta}|^4+\frac{b_2}{2}\left\{\eta_x^{* 2} \eta_y^2+\eta_x^2 \eta_y^{* 2}\right\} \\
& +b_3\left|\eta_x\right|^2\left|\eta_y\right|^2 +K_1\left\{\left|\partial_x \eta_x\right|^2+\left|\partial_y \eta_y\right|^2\right\} \\
& +K_2\left\{\left|\partial_x \eta_y\right|^2+\left|\partial_y \eta_x\right|^2\right\} +K_3\left\{\left(\partial_x \eta_x\right)^*\left(\partial_y \eta_y\right)+ c.c .\right\} \\
& +K_4\left\{\left(\partial_x \eta_y\right)^*\left(\partial_y \eta_x\right)+ c.c. \right\}\}.
\end{aligned}
\label{eq:freepwave}
\end{equation}
If we choose the GL parameters to be $b_3>|b_2|$ \cite{sigrist1991phenomenological}, the ground state is a nematic $p$-wave superconductor with the order parameter $ \vec{\eta} =(\eta_x,\eta_y) = (\eta,0)$, which breaks rotation symmetry reducing $D_{4h}$ (tetragonal) to $D_{2h}$ (orthorhombic). And the GL free energy density becomes
\begin{equation}
    F = \int d \boldsymbol{r}\{a|\eta|^2 + b_1|\eta|^4 + K_1|\partial_x \eta|^2  + K_2|\partial_y \eta|^2\},
\end{equation}
which has exactly the same form as Eq.~(1) of \textbf{Main Text}. However, the anisotropic effective masses here $K_1\neq K_2$ are not inherited from the anisotropic Fermi velocity in the normal state, but originate from the anisotropic form factor of the nematic $p$-wave pairing.

\section{\bf{\uppercase\expandafter{III. Requirements for the DC bias-induced SDE and USC}}}

\begin{figure}
		\centering
		\includegraphics[width=0.8\linewidth]{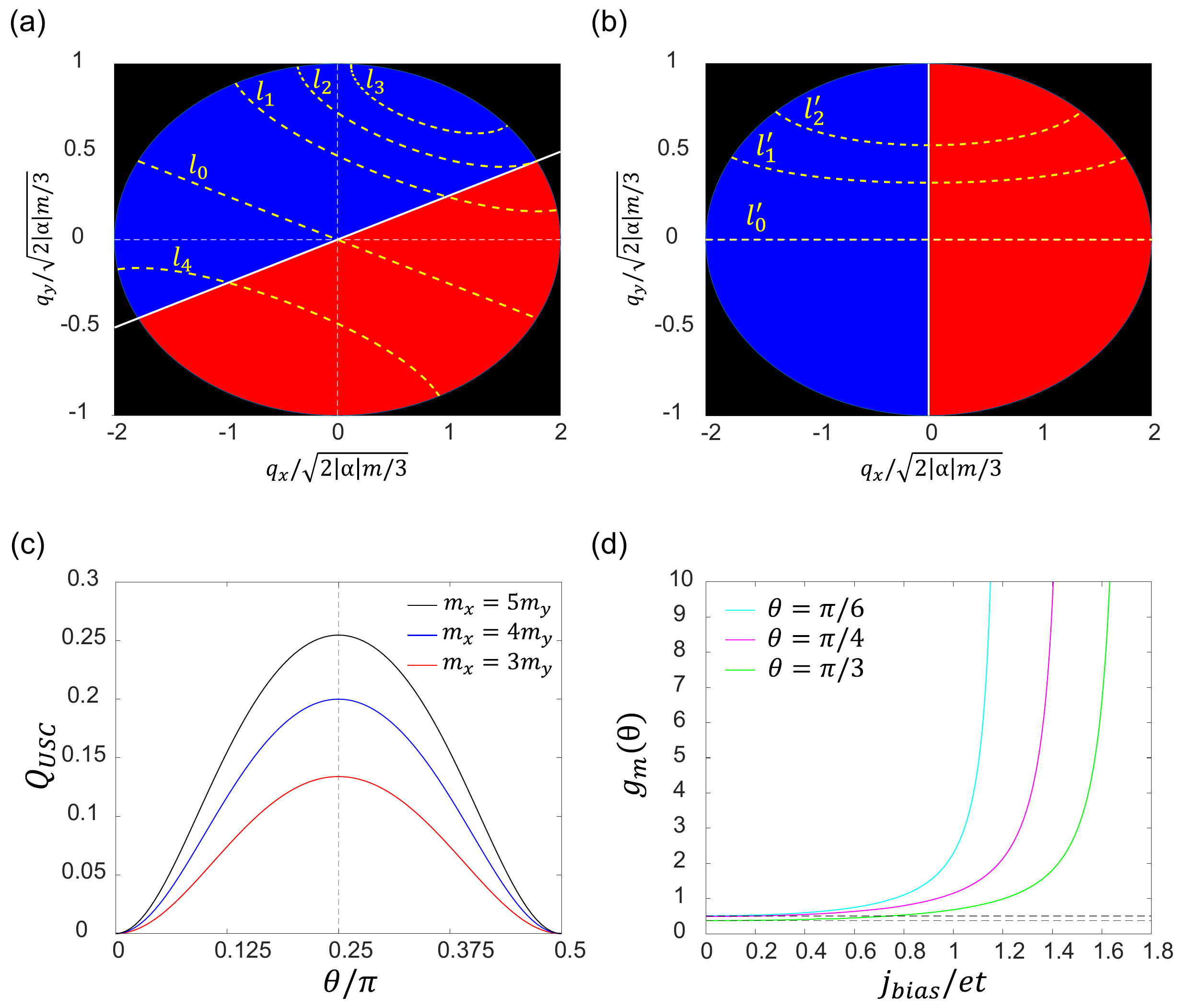}
		\caption{(a) The distribution of $j_{m}(q_x,q_y)$ (the black area is out of the domain), $j_{m} \leq 0$ inside the blue area, and $j_{m} \geq 0$  inside the red area. The yellow dotted curves $l_i$ ($i = 0$-$4$) show the curves determined by the equation: $j_{\parallel}(q_x,q_y) = j_{bias}$ with specific $j_{bias}$ ($
        j_{bias,0} = 0, j_{bias,1} = 0.9et, j_{bias,2} = 1.217et, j_{bias,3} = 1.4et$ and $ j_{bias,4} = -0.9et$). Parameter used here: $\theta = \pi/4, \alpha = -0.5t, \beta = 0.1t, m_x = 4/t, m_y = 1/t$. Then $j_{act} = j_{c+}(\theta) = 1.217et$. (b) The distribution of $j_{m}(q_x,q_y)$ when $\theta = \pi/2$ for comparison, the yellow dotted curves $l_{i}^{\prime}$ ($i = 0$-$2$) show the curves determined by the equation: $j_{\parallel}(q_x,q_y) = j_{bias}$ with specific $j_{bias}$ ($
        j_{bias,0}^{\prime} = 0, j_{bias,1}^{\prime} = 0.9et, j_{bias,2}^{\prime} = 1.4et$.) (c) The quality factor $Q_{USC}(\theta)$ of different effective mass $m_x$ and $m_y$, $Q(\theta)$ always has a local maximum at $\theta = \pi/4$. (d) The evolution of differential gain $g_m(\theta)$ with increasing bias current $j_{bias}$ for three different $\theta$, $g_m(\theta)$ is a constant in small $j_{bias}$ limit.}
		\label{fig:supfig1}
\end{figure}

In this section, we consider the conditions for the DC bias-induced SDE and USC. When the bias is injected in the direction $\hat{\theta}=(\cos\theta,\sin\theta)$, let $l$ be the curve determined by the equation: $j_{\parallel}(q_x,q_y) = j_{bias}$. A current carrying state with $j_{bias}$ should satisfy $(q_x, q_y)\in l$ and $(q_x, q_y) \in \mathcal{D}$ at the same time, where $\mathcal{D}$ is the domain of all physical $(q_x, q_y)$. Then we consider the measured transverse supercurrent $j_{m}$, we note that $j_{m} \leq 0$ is equivalent to
\begin{equation}
	\alpha_x^{\prime} \sin\theta \leq \alpha_y^{\prime} \cos\theta.
    \label{app:boundary2}
\end{equation}
In the context of the \textbf{Main Text}, we have $\frac{q_x\sin\theta }{m_x} \leq \frac{q_y\cos\theta }{m_y}$, and then
\begin{equation}
	q_y \geq \left( \frac{ m_y}{ m_x} \tan\theta \right) q_x  = k_{\theta} q_x,
    \label{main:boundary2}
\end{equation}
where $k_{\theta} = \frac{ m_y}{ m_x} \tan\theta$.
As shown in Fig.~\ref{fig:supfig1}(a), the line $q_y = k_{\theta} q_x$ (white line) divides the domain into a positive region and a negative region of $j_m$, within the red area and the blue area, respectively. Let $l_i$ be the curve determined by the equation: $j_{\parallel}(q_x,q_y) = j_{bias,i}$. If some of the points on $l_i$ fall into one side of the region, some fall into another side, but overall $l_i$ is not symmetric with respect to $q_y = k_{\theta} q_x$, we say the SDE in the transverse direction occurs. The physical picture here is that the control bias selects a finite-momentum condensate and effectively shifts the reference point for transverse current measurements. This geometric effect immediately produces $j_{mc+}\neq-j_{mc-}$—a transverse SDE emerging purely from anisotropy conversion. Further, if all the points on $l_i$ fall into one side of the region, we say the USC occurs, only the supercurrent flowing in one transverse direction is permitted.

Seeking a better understanding of this problem, we show the variation of the curves $l_i$ ($i = 0,1,2,3,4$) with various $j_{bias,i}$ and $\theta=\pi/4$ in Fig.~\ref{fig:supfig1}(a). $l_0$ corresponds to $j_{bias,0} = 0$. In this case, $l_0$ is symmetric with respect to the solid white line $q_y = k_{\theta} q_x$ and $j_{mc,0+}=-j_{mc,0-}$, i.e., no SDE. This is ensured by $\mathcal{T}$ and $\mathcal{P}$ symmetry. For $j_{c-}(\theta)<j_{bias,1} < j_{c+}(\theta)$, the curves $l_1$ crosses both the blue and red areas, so $j_{m}$ can be either negative or positive. However, $l_1$ is asymmetric with respect to the solid white line, thus SDE appears in this curve. The case for $l_4$ is similar to $l_1$ for $j_{bias,4}=-j_{bias,1}$. However, $l_4$ is a centrally symmetric mirror image of $l_1$, making $j_{mc,4\pm}=-j_{mc,1\mp}$. This is also the manifestation of the global $\mathcal{T}$ and $\mathcal{P}$ symmetry. $l_2$ satisfies $j_{bias,2} = j_{c+}(\theta)$, the solid white line $q_y = k_{\theta} q_x$ becomes the tangent of $l_2$ in this scenario, so this curve exhibits the ideal SDE. Finally, because of $j_{bias,3} > j_{c+}(\theta)$, $l_3$ is fully inside the blue area, which means $j_{m}$ can only be negative and shows USC. We then conclude that the minimal bias current needed to establish the USC is $j_{act} = j_{c+}(\theta)$, and the measured transverse current $j_{m}$ becomes unidirectional only when $j_{bias} \geq j_{c+}(\theta)$, which is consistent with the discussions in \textbf{Main Text}. As a comparison, we also plot curves $l_i^{\prime}$ ($i = 0,1,2$) with various $j_{bias,i}$ and $\theta=\pi/2$ in Fig.~\ref{fig:supfig1}(b). In this case, all $l_i^{\prime}$ are symmetric with respect to the $q_y$ axis, implying the vanishing of SDE when the bias is aligned with a high symmetry axis.

\section{\bf{\uppercase\expandafter {IV. Derivation of unidirectional superconductivity quality factor $Q_{USC}$, rectifier quality factor $Q_{rec}$ and differential gain $g_m$}}}

In this section, we want to derive $Q_{USC} = 1-j_{act}/j_{cut}$, $Q_{rec} = |{j_{mc1}}/{j_{act}}|$, and $g_m=\partial\eta/\partial j_{bias}$ at $j_{bias}\to0$. 

Since $j_{act} = j_{c+}(\theta)$, to obtain $Q_{USC}$ and $Q_{rec}$, here we only need to derive $j_{cut}$ and $j_{mc1}$. To determine $j_{cut}$, we note that the curve $j_{\parallel}(q_x,q_y) = j_{cut}$ and the domain boundary ${q_{x}^2}/{m_x} + {q_{y}^2}/{m_y} = {2|\alpha|}/{3}$ share a single intersection point, so if we combine these two equations, we can obtain the quadratic equation:
\begin{equation}
    \frac{m_x}{\cos^2\theta}\left( \frac{3\beta}{2e|\alpha|}j_{cut} - \frac{q_y\sin\theta }{m_y} \right)^2 + \frac{q_y^2}{m_y} = \frac{2|\alpha|}{3}.
\end{equation}
When this quadratic equation has only one solution, the discriminant must be 0, so we can obtain
\begin{equation}
    j_{cut} = \frac{e}{\beta}\left(\frac{2|\alpha|}{3}\right)^{3/2}\sqrt{\frac{\cos^2\theta}{m_x}+\frac{\sin^2\theta}{m_y}}.
\end{equation}
Then $q_y$ is given by:
\begin{equation}
    q_{y} = \sqrt{\frac{2|\alpha|}{3} } \sqrt{\frac{m_x m_y \sin^2 \theta}{m_x \sin ^2 \theta+m_y \cos ^2 \theta}},
\end{equation}
and we know $(q_x,q_y)$ is located at the boundary: ${q_{x}^2}/{m_x} + {q_{y}^2}/{m_y} = {2|\alpha|}/{3}$, so
\begin{equation}
    q_x = \sqrt{\frac{2|\alpha|}{3} } \sqrt{\frac{m_x m_y \cos^2 \theta}{m_x \sin ^2 \theta+m_y \cos ^2 \theta}}.
\end{equation}
From $j_{m}(q_x,q_y)$, $j_{mc2}$ is
\begin{equation}
    j_{mc2} = \frac{e}{\beta}\left(\frac{2|\alpha|}{3}\right)^{3/2} \frac{\left(m_y-m_x\right)\sin\theta \cos\theta}{\sqrt{\left(m_x \sin ^2 \theta+m_y \cos ^2 \theta\right) m_x m_y}}.
\end{equation}
In terms of $j_{act} = j_{c+}(\theta)$ and $j_{cut}$, we have
\begin{equation}
Q_{USC}=1 - \sqrt{\frac{m_xm_y}{(m_x\cos^2\theta+m_y\sin^2\theta)(m_y\cos^2\theta+m_x\sin^2\theta)}}. 
\end{equation}
And $Q_{USC}$ for different $(\theta,m_x,m_y)$ is plotted in Fig.~\ref{fig:supfig1}(c). It worth mention that that the maximum of $Q_{USC}(\theta)$ always locate at $\theta = \pi/4$, which is independent of the effective mass $m_x$ and $m_y$, and:
\begin{equation}
    \text{max}(Q_{USC}) = Q_{USC}(\pi/4) = 1-\frac{2\sqrt{m_xm_y}}{m_x+m_y}.
    \label{Q_USC_Max}
\end{equation}
We note that $\text{max}(Q_{USC})$ reveals a simple relationship to the degree of anisotropy of the superconductor. It is symmetric with respect to $m_x$ and $m_y$, and vanishes at the isotropic limit $m_x=m_y$. When $m_x/m_y<1$ or $m_x/m_y>1$, $\text{max}(Q_{USC})$ towards unity monotonically as $m_x/m_y\to0$ and $m_x/m_y\to\infty$, respectively.

To get $j_{mc1}$, we can solve the following equations simultaneously:
\begin{equation}
\left\{\begin{aligned}
& j_{act} = - \frac{e}{\beta} \left[ \alpha + \frac{1}{2}\left( \frac{q_x^2}{m_x} + \frac{q_y^2}{m_y} \right) \right]  \left( \frac{q_x\cos\theta }{m_x} + \frac{ q_y\sin\theta}{m_y} \right)  \\
& \frac{q_{x}^2}{m_x} + \frac{q_{y}^2}{m_y} =\frac{2|\alpha|}{3} \\
\end{aligned}
\label{app:setofeq2}
\right. ,
\end{equation}
one solution of $(q_x,q_y)$ corresponds to the case $j_{m}=0$, and the other solution is
\begin{equation}
\left\{\begin{aligned}
        q_{y} & = \frac{m_x+\left( m_x-m_y\right)\cos^2\theta}{m_x\sin^2\theta+m_y\cos^2\theta} \frac{\sqrt{2|\alpha|}m_y\sin\theta}{\sqrt{3(m_x\cos^2\theta+m_y\sin^2\theta)}} \\
        q_{x} & = \frac{m_x}{\cos\theta}\left( \frac{3\beta}{2e|\alpha|}j_{act}-\frac{q_y \sin\theta}{m_y} \right) \\
\end{aligned},\right. 
\label{solution2}
\end{equation}
and then $j_{mc1}$ can be determined as:
\begin{equation}
    \begin{aligned}
            & j_{mc1} = \frac{4e|\alpha|(m_y-m_x)}{3\beta} \sqrt{\frac{2|\alpha|}{3(m_x\cos^2\theta+m_y\sin^2\theta)}}\frac{\tan\theta}{m_x\tan^2\theta+m_y},
    \end{aligned}
\end{equation}
at the same time, we can obtain
\begin{equation}
    \begin{aligned}
        Q_{rec}(\theta) & = \left|\frac{j_{mc1}}{j_{act}}\right| = \frac{2|m_x-m_y|\tan\theta}{m_x\tan^2\theta+m_y}.
    \end{aligned}
\end{equation}

\begin{figure}[t]
		\centering
		\includegraphics[width=0.8\linewidth]{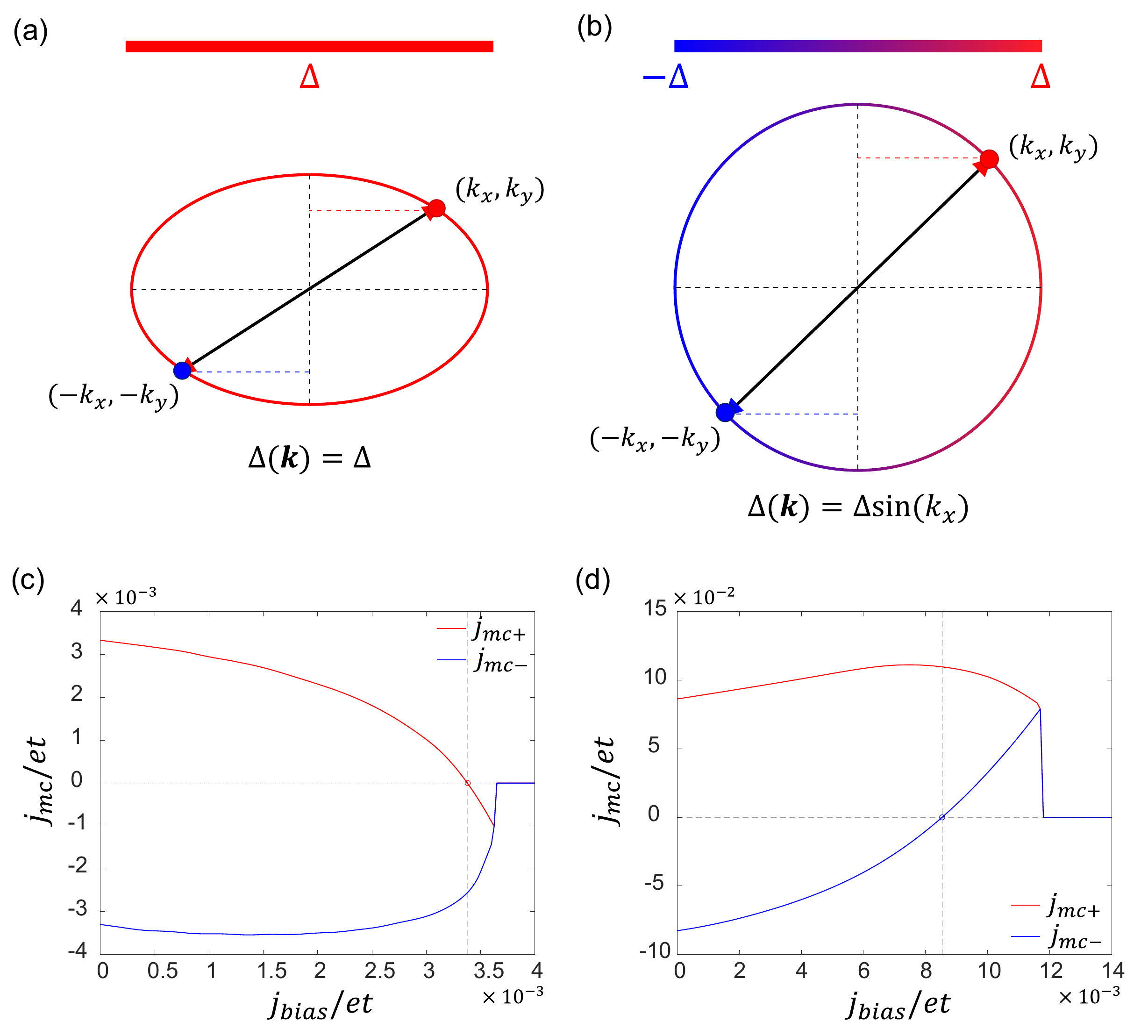}
		\caption{(a) Schematic illustration of anisotropic Fermi surface with $s$-wave pairing. (b) Schematic illustration of an isotropic Fermi surface with nematic $p$-wave pairing.  (c) The results of the square lattice model ($t_x \neq t_y$) with $s$-wave pairing, $j_{mc+}$ becomes negative when $j_{bias} > j_{act} = 0.0034et$. Parameter used here: $\theta = \pi/4, t_x = 0.5t_y = t$, $\mu = 2.5t$, $U = 1.5t$.
        (d) The results of the square lattice model ($t_x = t_y$) with $p$-wave pairing, $j_{mc-}$ becomes positive when $j_{bias} > j_{act} = 0.0086et$. Parameter used here: $\theta = \pi/4, t_x = t_y = t$, $\mu = 2.5t$, $U = 3.25t$.   }
		\label{fig:supfig2}
\end{figure}

To get $g_m$, we firstly need to find the measured critical currents in the transverse direction $j_{mc\pm}$, here we still employ Eq.~\eqref{app:setofeq2} but replacing $j_{act}$ by $j_{\parallel}=j_{bias}$ with small $j_{bias}$. In the small bias current limit (the higher order terms of $j_{bias}$ are omitted), we can get:
\begin{equation}
\left\{\begin{aligned}
        q_{yc} & \approx \pm \sqrt{\frac{2|\alpha|m_y^2}{3\left(m_x \tan ^2 \theta+m_y\right)}}\\&+\frac{3 \beta m_x m_y \sin \theta j_{b i a s}}{2 e|\alpha| \left(m_x \sin^2 \theta+ m_y\cos^2\theta\right)}  \\
        q_{xc} & \approx \frac{m_x}{\cos\theta}\left( \frac{3\beta}{2e|\alpha|}j_{bias}-\frac{q_y \sin\theta}{m_y} \right) \\
\end{aligned},\right. 
\label{solution1}
\end{equation}
And by substituting Eq.~\eqref{solution1} into $j_{m}(\bm{q})$, the transverse current $j_{mc\pm}$ is given by:
\begin{equation}
\begin{aligned}
j_{mc\pm}&=\pm \frac{2 e|\alpha|}{3 \beta \cos \theta} \sqrt{\frac{2|\alpha|}{3\left(m_x \tan ^2 \theta+m_y\right)}}\\&+\frac{ j_{bias} \left(m_y-m_x\right)}{m_x \tan  \theta+m_y\cot \theta}.
\end{aligned}
\label{jmeasurec}
\end{equation}
Then we can use $j_{mc\pm}$ to calculate the diode efficiency $\eta = \left|j_{mc+}+j_{mc-}|/|j_{mc+}-j_{mc-}\right|$. For small $j_{bias}$ limit, $\eta$ is approximately linear in $j_{bias}$ as $\eta\approx g_m(\theta)j_{bias}$, and $g_m(\theta)$ reads
\begin{equation}
g_m(\theta) =  \frac{\beta}{e}\left(\frac{3}{2|\alpha|}\right)^{3/2}\frac{|(m_y-m_x)\sin\theta\cos\theta|}{\sqrt{m_x\sin^2\theta+m_y\cos^2\theta}}.
\label{efficiency_small_limit}
\end{equation}
$g_m(j_{bias})$ for different $\theta$ is plotted in Fig.~\ref{fig:supfig1}(d). As we explained in \textbf{Main Text}, $g_m(\theta)$ is nearly a constant for small $j_{bias}<j_{act}$, but increases rapidly when $j_{bias}>j_{act}$, implying large differential gain in the USC region of CGOST.

\section{\bf{\uppercase\expandafter {V. Formalism and Numerical results of self-consistent mean-field calculations}}}

 In this section, we introduce the self-consistent mean-field (MF) theory~\cite{daido2022intrinsic,daido2022superconducting,sun2024flat} used in the \textbf{Main Text}. The total Hamiltonian is:
\begin{equation}
    \hat{H} = \hat{H}_0 + \hat{H}_{\mathrm{int}},
\end{equation}
where $\hat{H}_0 = \sum_{\bm{k}\sigma} \xi_n(\bm{k}) c_{\bm{k}\sigma}^\dagger c_{\bm{k}\sigma}$ and 
\begin{equation}
\hat{H}_{\mathrm{int}}=-\sum_{\bm{k}\bm{k}^{\prime}\bm{q}^{\prime}\sigma\sigma^{\prime}}U_{\sigma\sigma^{\prime}}(\bm{k},\bm{k}^{\prime},\bm{q}^{\prime})\hat{c}_{\bm{k}+\bm{q^{\prime}}\sigma}^{\dagger}\hat{c}_{-\bm{k}\sigma^{\prime}}^{\dagger}\hat{c}_{-\bm{k}^{\prime}\sigma^{\prime}}\hat{c}_{\bm{k}^{\prime}+\bm{q^{\prime}}\sigma}.
\end{equation}

The attractive interaction is \cite{daido2022superconducting}:
\begin{equation}
    U_{\sigma\sigma^{\prime}} (\bm{k},\bm{k}^{\prime},\bm{q}^{\prime}) = U_{\sigma\sigma^{\prime}}\delta_{\bm{q}\bm{q}^{\prime}}u_{\sigma\sigma^{\prime}}(\bm{k})u_{\sigma\sigma^{\prime}}^{*}(\bm{k}^{\prime}),
\end{equation}
where $\bm{q}$ is a finite center-of-mass momentum and $u_{\sigma\sigma^{\prime}}$ is the form factor.

To describe the current-flowing state, the superconducting order parameter $\Delta_{\sigma\sigma^{\prime}}(\bm{q})$ is considered with a center-of-mass momentum $\bm{q}=q_x \hat{x} + q_y \hat{y}$. The mean-field decoupling of the attractive interaction yields the following interaction Hamiltonian:
\begin{equation}
\begin{aligned}
    \hat{H}_{\mathrm{int}} &  = \sum_{\bm{k}\sigma\sigma^{\prime}} \Delta_{\sigma\sigma^{\prime}}(\bm{q}) u_{\sigma\sigma^{\prime}}(\bm{k}) c_{\bm{k}+\bm{q} \sigma}^{\dagger} c_{-\bm{k} \sigma^{\prime}}^{\dagger}  \\
    & + h.c. + \sum_{\sigma\sigma^{\prime}}\frac{N_c|\Delta_{\sigma\sigma^{\prime}}(\bm{q})|^2}{U_{\sigma\sigma^{\prime}}}.
\end{aligned}
\end{equation}
The full mean-field Hamiltonian can be rewritten in the Nambu basis $\psi(\bm{k},\bm{q}) =\left(\hat{c}_{\bm{k}+\bm{q}\uparrow},\hat{c}_{\bm{k}+\bm{q}\downarrow},\hat{c}_{-\bm{k}\uparrow}^{\dagger},\hat{c}_{-\bm{k}\downarrow}^{\dagger}\right)^T$ as
\begin{equation}
\begin{aligned}
    \hat{\mathcal{H}}_{\mathrm{MF}} 
    & =\sum_{\bm{k}}\psi^{\dagger}(\bm{k},\bm{q})\hat{H}_{\mathrm{BdG}}(\bm{k},\bm{q})\psi(\bm{k},\bm{q}) \\
    & + \sum_{\bm{k}}\operatorname{Tr}[\hat{H}_{n}(-\bm{k})] + \sum_{\sigma\sigma^{\prime}}\frac{N_c|\Delta_{\sigma\sigma^{\prime}}(\bm{q})|^2}{U_{\sigma\sigma^{\prime}}},
\end{aligned}
\end{equation}
where the Bogoliubov–de Gennes (BdG) Hamiltonian $\hat{H}_{\mathrm{BdG}}(\bm{k},\bm{q})$ reads:
\begin{equation}
\hat{H}_{\mathrm{BdG}}(\bm{k},\bm{q}) = \left(\begin{array}{cc}
         \hat{H}_n(\bm{k}+\bm{q})  & \hat{\Delta}(\bm{k},\bm{q}) \\
        \hat{\Delta}^{\dagger}(\bm{k},\bm{q}) & 
         -\hat{H}_n(-\bm{k}) 
    \end{array}
    \right),
\end{equation}
with $\hat{H}_n(\bm{k}) = \xi_n(\bm{k})\textbf{I}_{2}$ and the order parameter matrix is defined as:
\begin{equation}
\hat{\Delta}(\bm{k},\bm{q}) = \left(\begin{array}{cc}
         \Delta_{\uparrow\uparrow}(\bm{q})u_{\uparrow\uparrow}(\bm{k})  & \Delta_{\uparrow\downarrow}(\bm{q})u_{\uparrow\downarrow}(\bm{k}) \\
        -\Delta_{\uparrow\downarrow}(\bm{q})u_{\uparrow\downarrow}(\bm{k}) & 
         \Delta_{\downarrow\downarrow}(\bm{q})u_{\downarrow\downarrow}(\bm{k}) 
    \end{array}
    \right).
\end{equation}

And the grand potential can be defined as: 
\begin{equation}
    \Omega[\mu(\bm{q}),\Delta(\bm{q}),\bm{q}] = -\frac{1}{\beta}\operatorname{log}\mathcal{Z} = - \frac{1}{\beta}\operatorname{ln}\operatorname{Tr}[e^{-\beta\hat{\mathcal{H}}_{\mathrm{MF}}}],
\end{equation}
In the canonical ensemble, we transfer to the Helmholtz free energy density $\mathcal{F}$, which is written as:
\begin{equation}
\begin{aligned}
    & \mathcal{F}(\bm{q}) = \frac{1}{N_c}\left(\Omega[\mu(\bm{q}),\Delta(\bm{q}),\bm{q}]+\mu(\bm{q})N_c\right) \\
    & = -\frac{1}{\beta N_c} \sum_{n \bm{k}} \ln \left(1+e^{-\beta E_n(\bm{k}, \bm{q})}\right) +\sum_{\sigma\sigma^{\prime}}\frac{|\Delta_{\sigma\sigma^{\prime}}(\bm{q})|^2}{U_{\sigma\sigma^{\prime}}}.
\end{aligned}
\label{helmholzF}
\end{equation}
where $E_n(\bm{k},\bm{q})$ is the n-th eigenvalue of the BdG Hamiltonian $\hat{H}_{\mathrm{BdG}}(\bm{k},\bm{q})$.
For a given $\bm{q}$, $\Delta(\bm{q})$ can be determined self-consistently by solving the gap equation:
\begin{equation}
    \Delta_{\sigma\sigma^{\prime}}(\bm{q})=-\frac{U_{\sigma\sigma^{\prime}}}{N_c} \sum_{\bm{k}}u^{*}_{\sigma\sigma^{\prime}}(\bm{k})\left\langle\hat{c}_{-\bm{k} \sigma^{\prime}} \hat{c}_{ \bm{k}+\bm{q} \sigma}\right\rangle.
\end{equation}
Then we can obtain the minimized Helmholtz free energy density $\mathcal{F}_m(\bm{q})$ by substituting solved $\hat{\Delta}(\bm{k},\bm{q})$ in Eq.~\eqref{helmholzF} and the supercurrent can be evaluated by $\vec{j}(\bm{q}) = 2e \nabla_{\bm{q}} \mathcal{F}_m(\bm{q})$.

In the main text, we only consider the interaction between electrons with opposite spins, so the explicit forms of $U_{\sigma\sigma^{\prime}}(\bm{k},\bm{k}^{\prime},\bm{q}^{\prime})$ can be simplified as follows:
\begin{equation}
    U_{\sigma\sigma^{\prime}}(\bm{k},\bm{k}^{\prime},\bm{q}^{\prime}) = U(1-\delta_{\sigma\sigma^{\prime}})\delta_{\bm{q}\bm{q}^{\prime}}u(\bm{k})u^{*}(\bm{k}^{\prime}),
    \label{Usinglet}
\end{equation}
where
\begin{equation}
u(\bm{k}) = \left\{\begin{aligned}
& 1 & \text{($s$-wave)} \\
& \sin k_x & \text{($p$-wave)} \\
& \cos k_x-\cos k_y & \text{($d$-wave)} \\
\end{aligned}
\right. .
\end{equation}
The BdG Hamiltonian can be reduced to a 2 by 2 matrix as:
\begin{equation}
\hat{H}_{\mathrm{BdG}}(\bm{k},\bm{q}) = \left(\begin{array}{cc}
         \xi_n(\bm{k}+\bm{q})  & {\Delta}(\bm{q})u(\bm{k}) \\
        \Delta^{*}(\bm{q})u^{*}(\bm{k}) & 
         -\xi_n(-\bm{k}) 
    \end{array}
    \right),
\end{equation}
and the gap equation reads:
\begin{equation}
    \Delta(\bm{q})=-\frac{U}{N_c} \sum_{\bm{k}}u^{*}(\bm{k})\left\langle\hat{c}_{-\bm{k} \downarrow} \hat{c}_{ \bm{k}+\bm{q} \uparrow}\right\rangle.
\end{equation}

To further support the analysis based on our phenomenological theory, in this section, we summarize the results for numerical calculations using self-consistent MF theory for the toy models introduced in \textbf{Main Text}. The schematics for the two cases are shown in Fig.~\ref{fig:supfig2}(a) and (b). The numerical results are summarized in Fig.~\ref{fig:supfig2}(c) and (d), where the curves follow a trend similar to those in Fig.~2(a) in \textbf{Main Text}. The SDE can occur as long as $j_{bias}>0$, and the USC can be realized when $j_{bias} \geq j_{c+}(\theta) = 0.0034et$ in (c) and $j_{c+}(\theta) = 0.0086et$ in (d). Compared with phenomenological analysis, we numerically verified that the anisotropies of both the Fermi velocity and the pairing symmetry can lead to DC bias-controlled SDE and USC. 

A more quantitative verification can be done by comparing $Q_{USC}(\pi/4)$. We recall that
\begin{equation}
    \text{max}(Q_{USC}) = Q_{USC}(\pi/4) = 1-\frac{2\sqrt{m_xm_y}}{m_x+m_y}.
    \label{Q_USC_Max}
\end{equation}

As we mentioned in \textbf{Main Text}, the $m_x/m_y$ in Eq.~\eqref{Q_USC_Max} can be numerically extracted from $m_x/m_y=\left(j_{c\pm}(\pi/2)/j_{c\pm}(0)\right)^2$. Thus, we can compare Eq.~\eqref{Q_USC_Max} (we denote as $Q_{USC,a}$) with the $Q_{USC,n}$ from direct numerical evaluation. For the $s$-wave pairing case, the numerical result $Q_{USC,n}$ matches perfectly with the analytical result $Q_{USC,a}$ ($Q_{USC,n} =  6.90\%$ and $Q_{USC,a} = 6.91\%$). For the $p$-wave pairing case, $Q_{USC,n} = 25.8\%$ and $Q_{USC,a} = 20.4\%$, which are basically in agreement.

\end{document}